\documentclass[twocolumn,superscriptaddress,nofootinbib]{revtex4-1}
\usepackage[utf8]{inputenc} 
\usepackage[T1]{fontenc}    
\usepackage{hyperref}       
\usepackage{url}            
\usepackage{booktabs}       
\usepackage{amsfonts}       
\usepackage{nicefrac}       
\usepackage{microtype}      
\usepackage{lipsum}
\usepackage{amsmath}
\usepackage{graphicx}
\usepackage{comment}
\usepackage{xcolor}         

\definecolor{midgreen}{rgb}{0.52, 0.73, 0.4}

\usepackage{float}

\newcommand{\Lagr}{\mathcal{L}}

\hypersetup{pdftex,colorlinks=true,linkcolor=blue,citecolor=magenta,menucolor=black,urlcolor=blue,filecolor=blue}

\begin{document}

%
%
\title{ QCD Equilibrium and Dynamical Properties from Holographic Black Holes}

\author{Joaquin Grefa}
\affiliation{Physics Department, University of Houston, Houston TX 77204, USA}
\author{Mauricio Hippert}
\affiliation{Illinois Center for Advanced Studies of the Universe, Department of Physics, University of Illinois at Urbana-Champaign, Urbana, IL 61801, USA}

\author{Jorge Noronha} 
\affiliation{Illinois Center for Advanced Studies of the Universe, Department of Physics, University of Illinois at Urbana-Champaign, Urbana, IL 61801, USA}

\author{Jacquelyn Noronha-Hostler}
\affiliation{Illinois Center for Advanced Studies of the Universe, Department of Physics, University of Illinois at Urbana-Champaign, Urbana, IL 61801, USA}

\author{Israel Portillo}
\affiliation{Physics Department, University of Houston, Houston TX 77204, USA}

\author{Claudia Ratti}
\affiliation{Physics Department, University of Houston, Houston TX 77204, USA}

\author{Romulo Rougemont}
\affiliation{Instituto de F\'{i}sica, Universidade Federal de Goi\'{a}s, Av. Esperan\c{c}a - Campus Samambaia, CEP 74690-900, Goi\^{a}nia, Goi\'{a}s, Brazil}

\date{\today}

\begin{abstract}

%
%
By using gravity/gauge correspondence, we employ an Einstein-Maxwell-Dilaton model to compute the equilibrium and out-of-equilibrium properties of a hot and baryon rich strongly coupled quark-gluon plasma. The family of 5-dimensional holographic black holes, which are constrained to mimic the lattice QCD equation of state at zero density, is used to investigate the temperature and baryon chemical potential dependence of the equation of state. We also obtained the baryon charge conductivity, and the bulk and shear viscosities with a particular focus on the behavior of these observables on top of the critical end point and the line of first order phase transition predicted by the model.

\end{abstract}

\maketitle

\section{Introduction}

It is the goal of heavy ion physics to map the phases of strongly interacting matter at finite temperature and density. Within the multidimensional QCD phase diagram, the most common representations is a plane of temperature and baryon chemical potential where the hadronic phase is located at low temperature and density, whereas at large temperature there is experimental evidence that the quarks and gluons behave similarly to a strongly interacting liquid that we now call the quark-gluon plasma (QGP) \cite{Shuryak:1980tp}.

It is well established that the transition between the hadron gas and the QGP is a smooth crossover at vanishing chemical potential \cite{Aoki:2006we,Bellwied:2015rza}. However, it is conjectured that the crossover must evolve into a line of first order phase transition with a critical end point (CEP) at some finite value of baryon chemical potential. In fact, this is the case in chiral models when the effects of finite quark masses are considered \cite{Stephanov:1998dy}. Experimentally, the QCD phase diagram is scanned by systematically decreasing the center of mass energy per nucleon ($\sqrt{s}$) in relativistic heavy ion collisions, thus favoring matter over anti-matter produced in these events \cite{STAR:2013gus}. On the theory side, lattice QCD simulations are the best non-perturbative tool available to study QCD thermodynamics and the transition from the hadronic phase to the deconfined QGP one, and in particular, lattice QCD has provided the equation of state (EoS) at zero chemical potential \cite{Aoki:2006we,Borsanyi:2016ksw}. Nevertheless, calculations at finite baryon density are limited by the sign problem, an obstacle present in any path integral representation of systems with fermions at finite density \cite{Philipsen:2012nu}. One way to circumvent this problem is extending the EoS via Taylor expansion, although this series procedure only works for moderate values of baryon chemical potential and is available up to a ratio of $\mu_{B}/T=3.5$ \cite{Borsanyi:2021sxv}. As a consequence, a large region in the QCD phase diagram remains unknown. 

Therefore, in order to potentially guide the experimental search of the QCD critical point, we need an effective field theory for hot and dense deconfined matter that exhibits near perfect fluidity, a feature of the QGP characterized by a surprisingly small value of its shear viscosity over entropy density ($\eta/s$) \cite{Heinz:2013th}. One must also require that this theoretical approach agrees with the lattice EoS and allows non-equilibrium calculations to study the dynamical response of the QGP encoded in the transport coefficients, which are not easily extracted from first principles despite the progress of ab initio lattice calculations \cite{Ratti_2018}. Besides the shear ($\eta$) and bulk viscosities ($\zeta$), used as inputs in the hydrodynamical simulations of evolution of relativistic heavy ion collisions \cite{Dore:2020jye}, the baryon diffusion coefficient and conductivity are also relevant at finite baryon density \cite{Denicol:2018wdp}.

The gauge/gravity holographic duality \cite{Maldacena:1997re,Gubser:1998bc,Witten:1998qj} can be employed to study strongly interacting matter at nonzero temperature and density both in equilibrium \cite{DeWolfe:2010he,Critelli:2017oub} and out of equilibrium \cite{DeWolfe:2011ts, Rougemont:2017tlu}. In this manuscript, we summarize the results regarding the equilibrium and dynamical properties of strongly interacting matter at finite density by using a family of 5-dimensional holographic black holes and its phase diagram from Refs. \cite{Grefa:2021qvt,Grefa:2022sav}.

\section{The Holographic EMD Model}

\begin{figure}[H]
    \centering
    \includegraphics[width=0.49\textwidth]{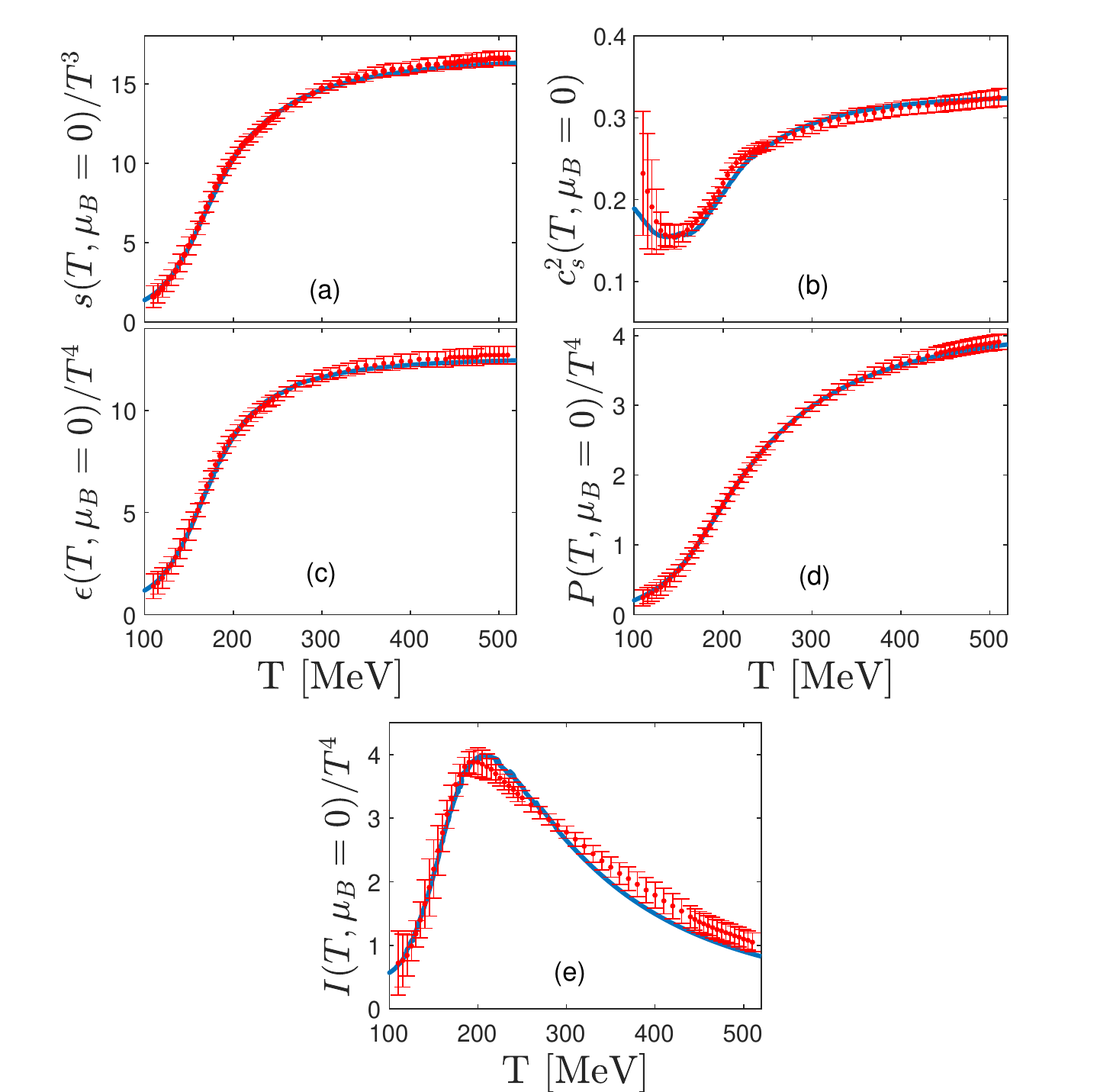}
    \caption{Thermodynamics at $\mu_{B}=0$. Lattice QCD results from Ref.\ \cite{Borsanyi:2013bia} (red points) are compared to the holographic model curves (blue lines): (a) entropy density, (b) speed of sound squared, (c) energy density $\epsilon$, (d) pressure $P$, and (e) trace anomaly $I=\epsilon-3P$.}
    \label{fig:latticeEoS_zero}
\end{figure}

The most general five-dimensional bulk action with one metric field, an abelian gauge field $A_{\mu}$, and a real scalar field $\phi$ with at most two derivatives that can generate a QCD-like theory \cite{DeWolfe:2010he} reads,
\begin{eqnarray} \label{eq:action}
    S&=&\int_{\mathcal{M}_5} d^5 x \Lagr = \frac{1}{2\kappa_{5}^{2}}\int_{\mathcal{M}_5} d^{5}x\sqrt{-g}\times
    \nonumber\\
    &\times&\left[R-\frac{(\partial_\mu \phi)^2}{2}-V(\phi)-\frac{f(\phi)F_{\mu\nu}^{2}}{4}\right],
\end{eqnarray}
where $k_{5}^{2}$ is the 5-dimensional gravitational constant, $g_{\mu\nu}$ is the metric tensor, $R$ is the Ricci scalar, $\phi$ is the dilaton field and $V(\phi)$ is a scalar potential that breaks conformal invariance in the quantum field theory side. $V(\phi)$ is a free function within the holographic model that is dynamically fixed by solving the equations of motion at $\mu_{B}=0$ with the constrain that the holographic equation of state matches the lattice results with (2+1) flavors and physical values of the quark masses at zero chemical potential from \cite{Borsanyi:2013bia}. The results of this procedure are shown in Fig. \ref{fig:latticeEoS_zero}.

\begin{figure}[H]
    \centering
    \includegraphics[width=0.5\textwidth]{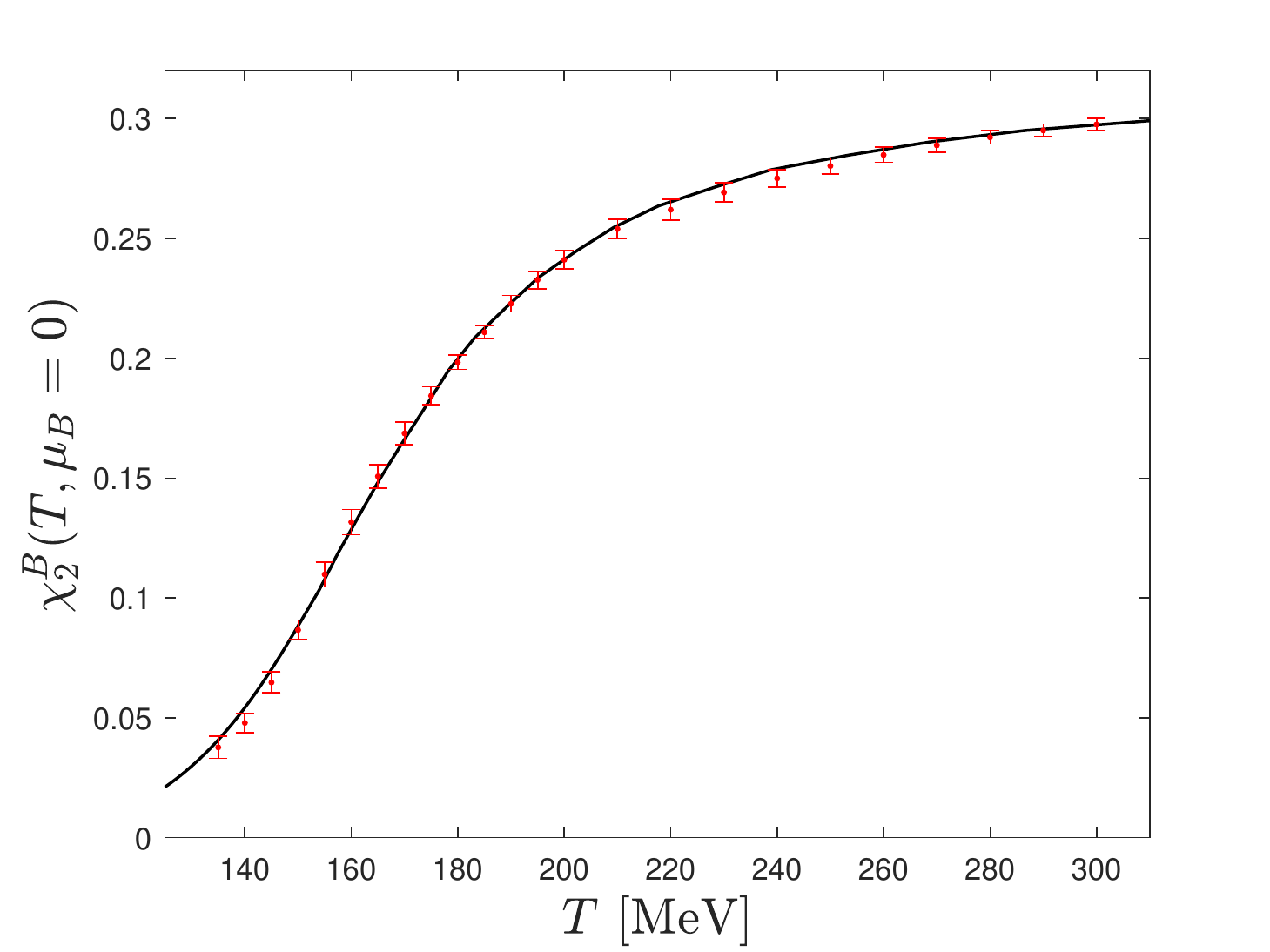}
    \caption{Results from the fitting of the holographic susceptibility (solid black curve) to the dimensionless second order baryon susceptibility $\chi_{2}^{B}(T,\mu_{B}=0)$ from lattice QCD \cite{Bellwied:2015lba}.}
    \label{fig:chi2_B0}
\end{figure}

 Additionally, the holographic approach allows to obtain an EoS at finite chemical potential by adding a Maxwell field $A_{\mu}$, and an additional function $f(\phi)$ that couples the dilaton and Maxwell fields, hence defining an Einstein-Maxwell-Dilaton model (EMD). The coupling function $f(\phi)$ is dynamically fixed by matching the solutions from the black hole to the second order baryon susceptibility at zero chemical potential obtained from lattice calculations as shown in Fig. \ref{fig:chi2_B0}. The $n-$th order baryon susceptibility is defined as:

\begin{equation}
    \chi_{n}=\frac{\partial^n (P/T^{4})}{\partial(\mu_{B}/T)^{n}},
\end{equation}
where $P$ corresponds to the pressure. The susceptibilities are the coefficients of the Taylor series expansion of the pressure and the baryon density $\rho_{B}$.

In the EMD model, the ansatz used to describe charged, isotropic, translational and rotationally invariant and asymptotically Anti-de Sitter (AdS) black hole solutions may be written as:

\begin{equation}
\begin{array}{rcl} \label{eq:ansatz}
     ds^2 &=& e^{2A(r)}[-h(r)dt^2+d\vec{x}^2]+\frac{dr^{2}}{h(r)}, \\
     \phi &=& \phi(r), \\
     A &=& A_{\mu}dx^{\mu}=\Phi(r)dt.
\end{array}
\end{equation}
with the boundary of the asymptotically AdS$_{5}$ spacetime placed at $r\rightarrow\infty$, the black hole horizon $r_{H}$ is given by the largest root of $h(r_{H})=0$ and the radius of the asymptotically AdS$_{5}$ background is set to unity.

From the action \ref{eq:action} and the Ansatz \ref{eq:ansatz}, the equations of motion (EOM) are numerically solved and the thermodynamics for a QCD-like theory is obtained through holographic dictionary, i.e. from the asymptotic behavior of the black hole background fields, one can compute the entropy density $s$ and the baryon density $\rho$ over a plane of temperature $T$ and baryon chemical potential $\mu_{B}$ as detailed in Ref. \cite{Critelli:2017oub,Grefa:2021qvt}.

It is important to point out that this holographic EMD model is fixed to mimic the lattice EoS only at zero chemical potential. Therefore, any calculation at finite $\mu_{B}$ and any calculations at $\mu_{B}=0$ concerning observables not employed to fix the free parameters of the model (e.g. the bulk viscosity at $\mu_{B}=0$ which is compared to Bayesian analyses in Ref. \cite{Grefa:2022sav}) constitute actual predictions of the model.

\section{The EoS and the Critical End Point}

\begin{figure}[H]
    \centering
    \includegraphics[width=0.5\textwidth]{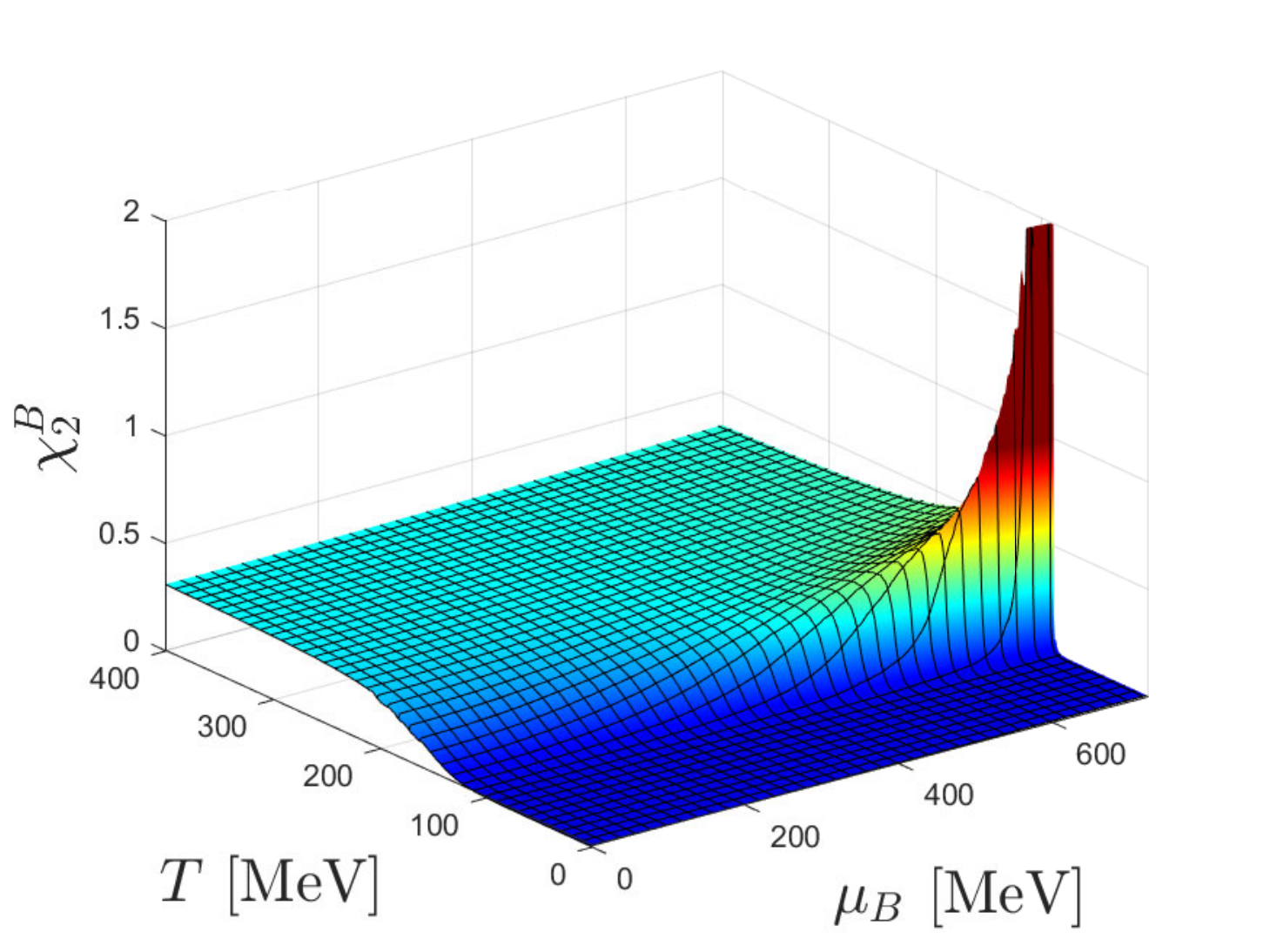}
    \includegraphics[width=0.5\textwidth]{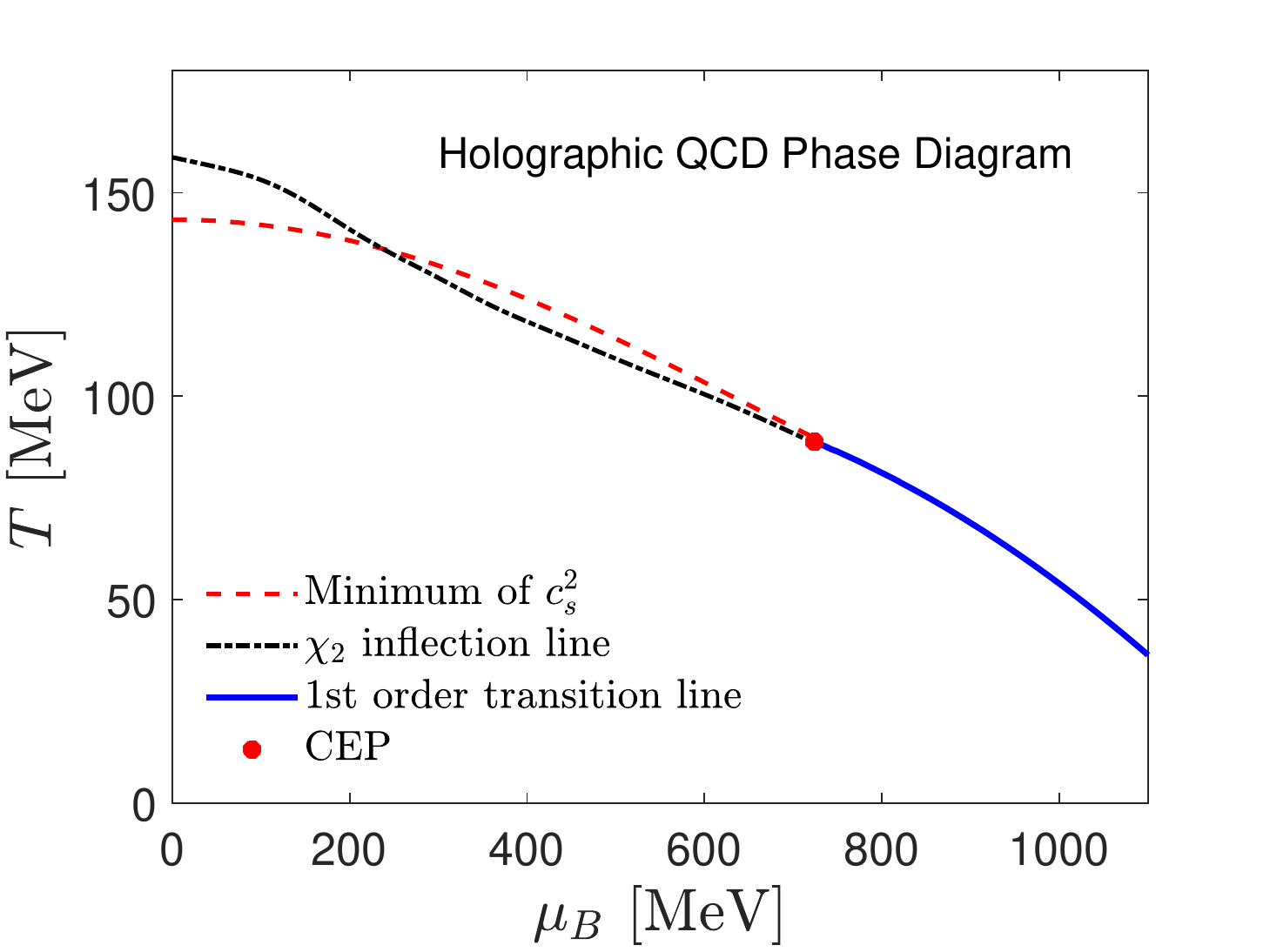}
    
    \caption{Upper panel: the behavior of the second order baryon susceptibility $\chi_{2}^{B}$ in the $(T,\mu_{B})$ plane. As the chemical potential increases, $\chi_{2}^{B}$ develops a peak that diverges at the critical point located at $T^{\textrm{CEP}}\sim 89$ MeV and $\mu_{B}^{\textrm{CEP}}\sim 724$ MeV. Lower panel: the phase diagram of our EMD model. The inflection point of $\chi_2^{B}$ and the minimum of the speed of sound squared at constant entropy per baryon number $c_{s}^{2}$ are used to characterize the crossover region.}
    \label{fig:phase_diagram_eq}
\end{figure}

Once the EoS for the QCD-like theory is obtained at finite chemical potential, it is possible to look for signatures of criticality. In fact, the second order baryon susceptibility $\chi_{2}^{B}$, which could be interpreted as a measure of how the baryon density reacts when $\mu_{B}$ is increased, exhibits a divergence as shown in Fig. \ref{fig:phase_diagram_eq}, at the coordinates  $T^{C}=89$ MeV and $\mu_{B}^{C}=724$ MeV, which is a candidate for the QCD critical end point (CEP).

Additionally, two state variables, sensitive the change of degrees of freedom from the confined hadronic phase to the high temperature deconfined QGP phase, were chosen to describe the crossover region. In particular, the inflection point of the second order baryon susceptibility $\chi_{2}^{B}$ and the minimum of the square of the speed of sound at constant entropy per baryon number $c_{s}^{2}$ were considered as pseudo transition lines. These trajectories follow and meet at the CEP.   

\begin{figure}[H]
    \centering
    \includegraphics[width=0.5\textwidth]{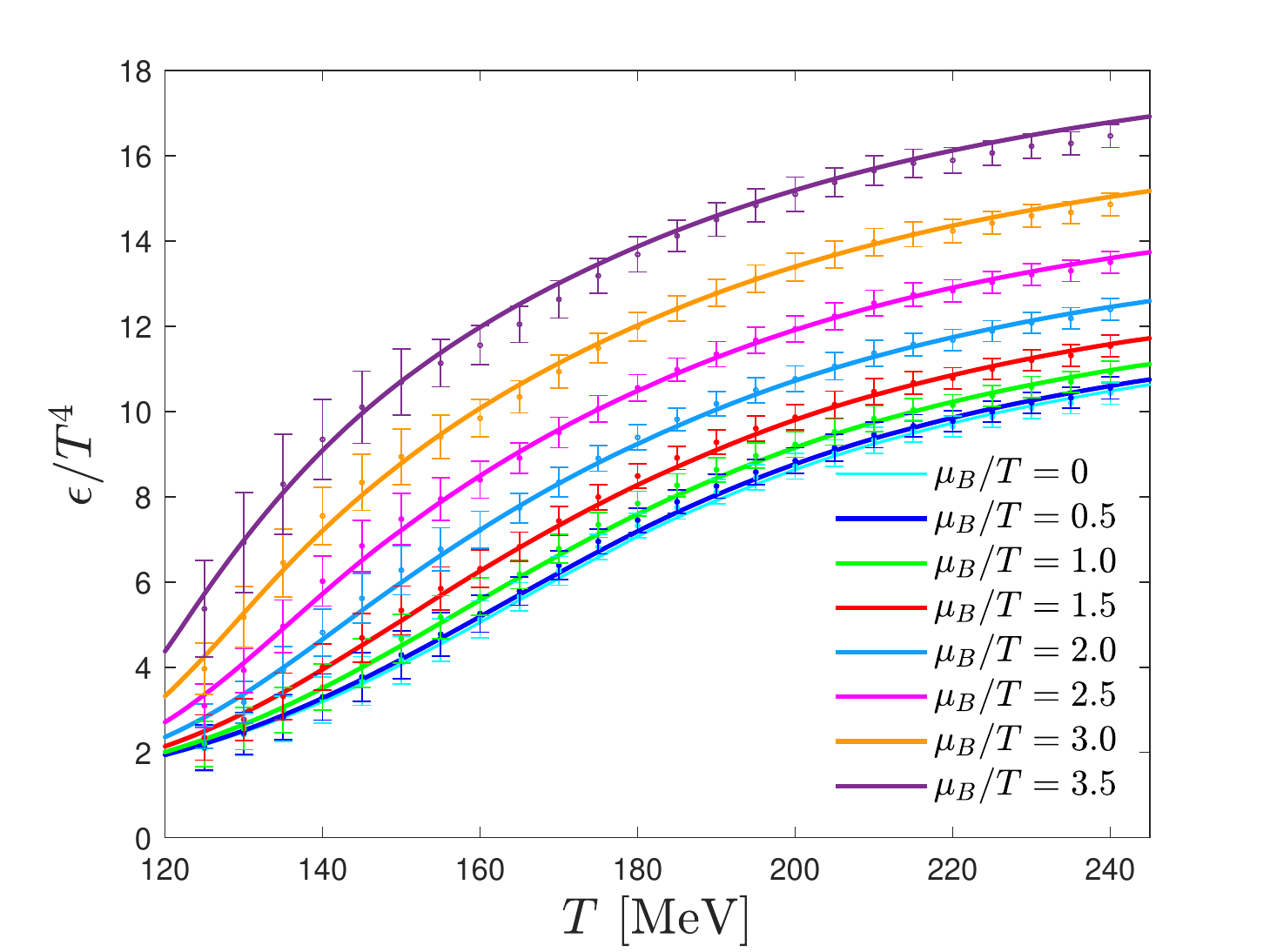}
    \caption{normalized energy density as a function of the temperature for different values of $\mu_{B}/T$ and its comparison with state-of-the-art lattice QCD results from Ref. \cite{Borsanyi:2021sxv}}
    \label{fig:energy_density}
\end{figure}

Thanks to numerical developments, presented in Ref. \cite{Grefa:2021qvt}, it is possible to obtain the thermodynamics of the QCD-like theory over a broad region in the phase diagram within the rectangle defined by $T \in [2,550]$ MeV and $\mu_B\in [0,1100]$ MeV. The holographic EoS is compared with the most recent lattice QCD results from an alternative expansion scheme \cite{Borsanyi:2021sxv}. This comparison between the holographic EMD and the state-of-the-art lattice results for the case of the energy density up to the unprecedented value of $\mu_{B}/T=3.5$ is shown in Fig. \ref{fig:energy_density} with an excellent quantitative agreement.

\section{Transport Coefficients}

The computation of the holographic transport coefficients, relevant to study the near-equilibrium response of the QGP, makes use of formulas that have been already derived in the literature. Here, we consider the black hole background fields solution and the holographic EoS from Refs. \cite{Critelli:2017oub,Grefa:2021qvt}. For the first time, we present the transport variables of hot and dense quark-gluon matter over the crossover region, on top of the critical point and across the line of first order phase transition as reported in \cite{Grefa:2022sav}.

In order to obtain the baryon conductivity, bulk viscosity, and shear viscosity, we consider linear disturbances of the finite temperature and baryon dense EMD black hole background fields. As a result of a plane wave Ansatz for the black hole background field perturbations with zero momentum and frequency $\omega$, the resulting disturbances can be organized into gauge and diffeomorphism invariant combinations of the $SO(3)$ symmetry group of spatial rotations \cite{DeWolfe:2011ts}. The $SO(3)$ singlet channel is holographically related to the bulk viscosity ($\zeta$) of the dual plasma, the triplet to the baryon conductivity ($\sigma_{B}$), and the quintuplet to the shear viscosity ($\eta$). 

\subsection{Baryon Conductivity}

\begin{figure}[H]
    \centering
    \includegraphics[width=0.5\textwidth]{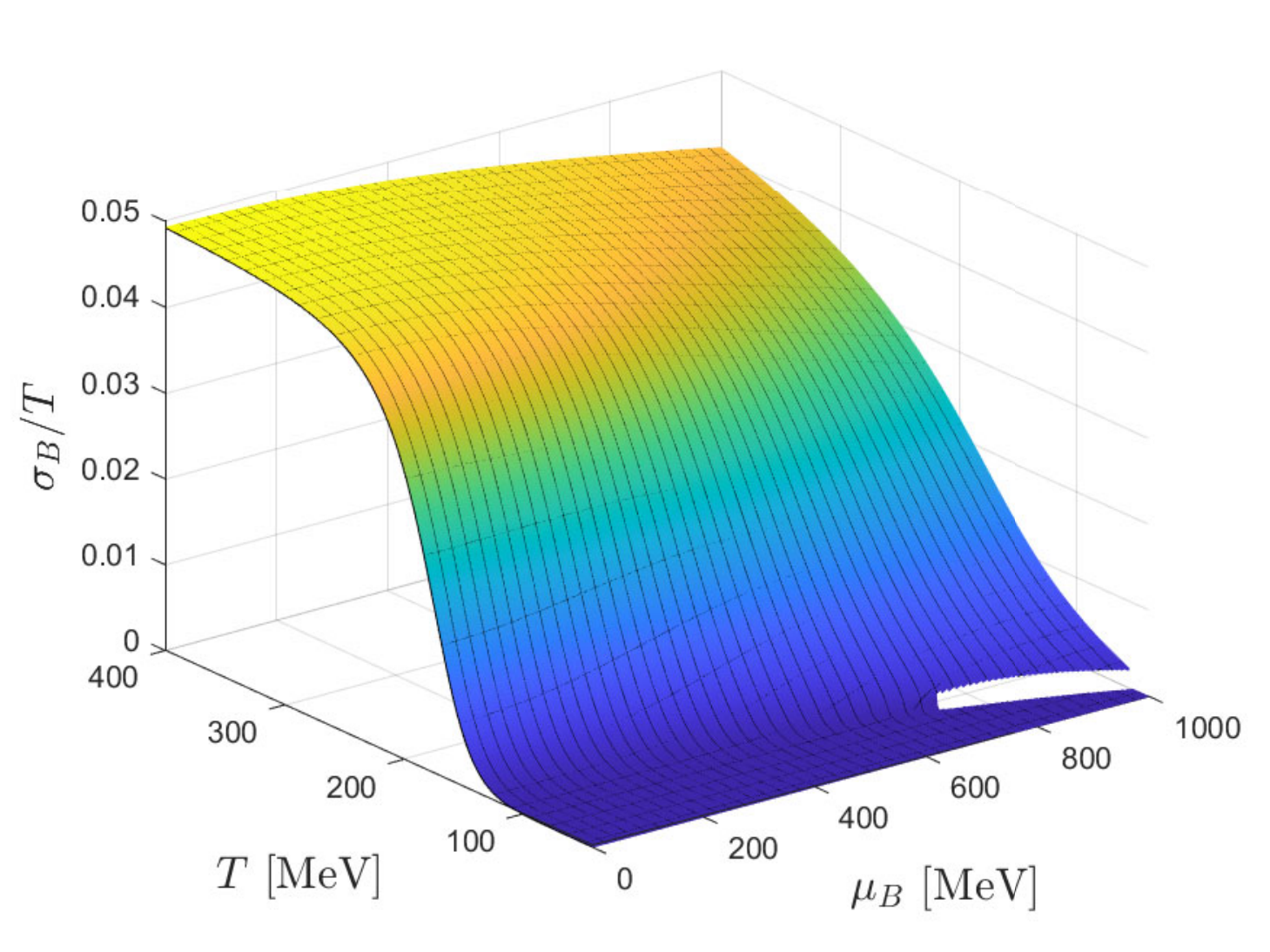}
    \includegraphics[width=0.5\textwidth]{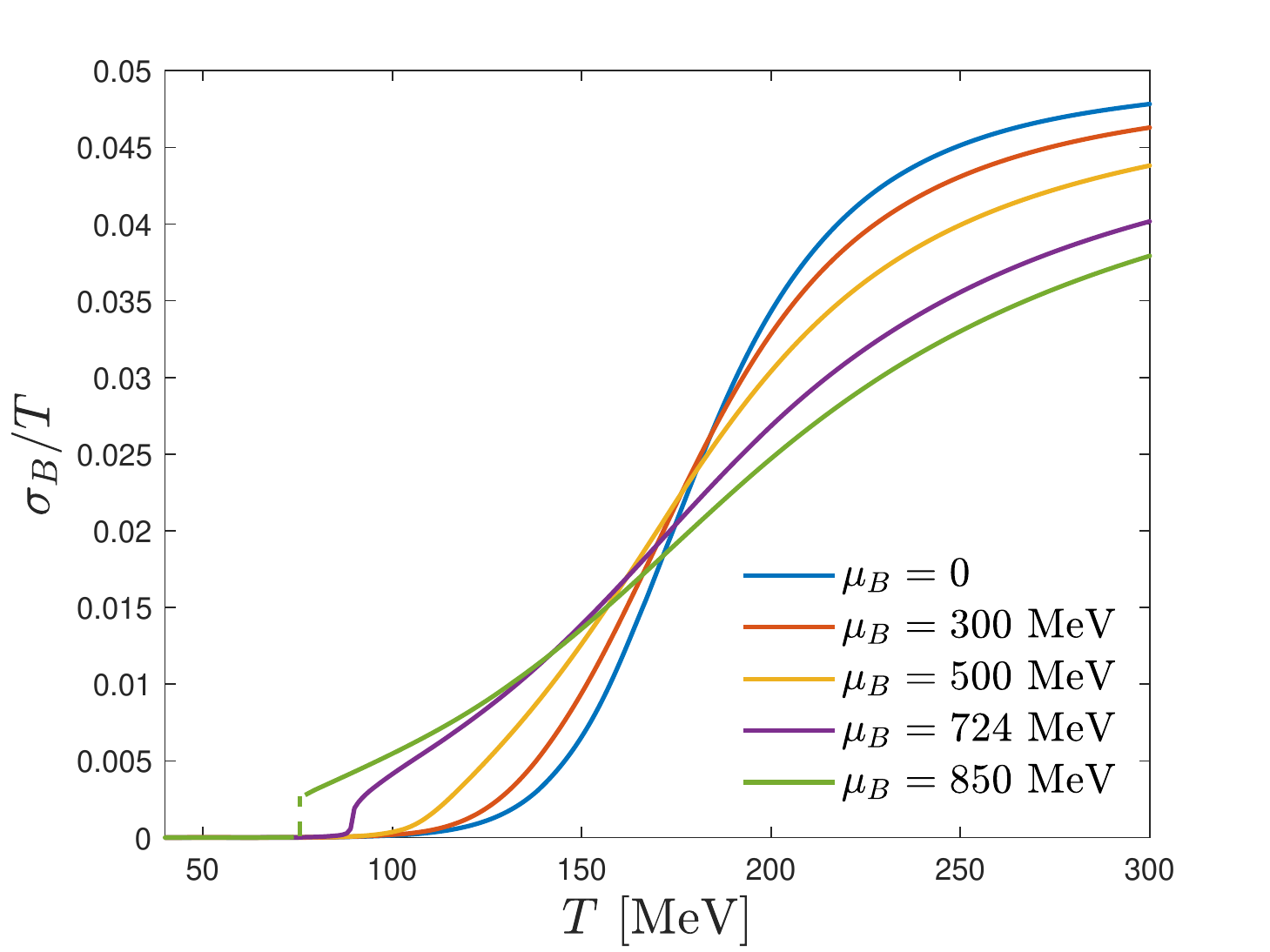}
    \caption{Upper panel: scaled baryon conductivity $\sigma_B/T$ as a function of temperature and baryon chemical potential. Lower panel: scaled baryon conductivity as a function of the temperature, for several values of the chemical potential.}
    \label{fig:barCond}
\end{figure}

Regarding the baryon conductivity $\sigma_{B}$, the derivation can be found in Ref. \cite{DeWolfe:2011ts}, the details of the numerical calculations are reported in Ref. \cite{Grefa:2022sav}, and the holographic results for the baryon conductivity are shown in Fig. \ref{fig:barCond}. Overall, the baryon conductivity does not strongly depend on the baryon chemical potential, and it remains finite at the location of the critical point, although it develops an infinite slope at this location. This indicates this holographic result for $\sigma_{B}$ lies in the type B dynamical universality class \cite{Hohenberg:1977ym}. Beyond the critical point and over the line of first order phase transition, $\sigma_{B}/T$ develops a discontinuity that remains small as the baryon chemical potential increases. 

\subsection{Bulk Viscosity}

\begin{figure}[H]
    \centering
    \includegraphics[width=0.5\textwidth]{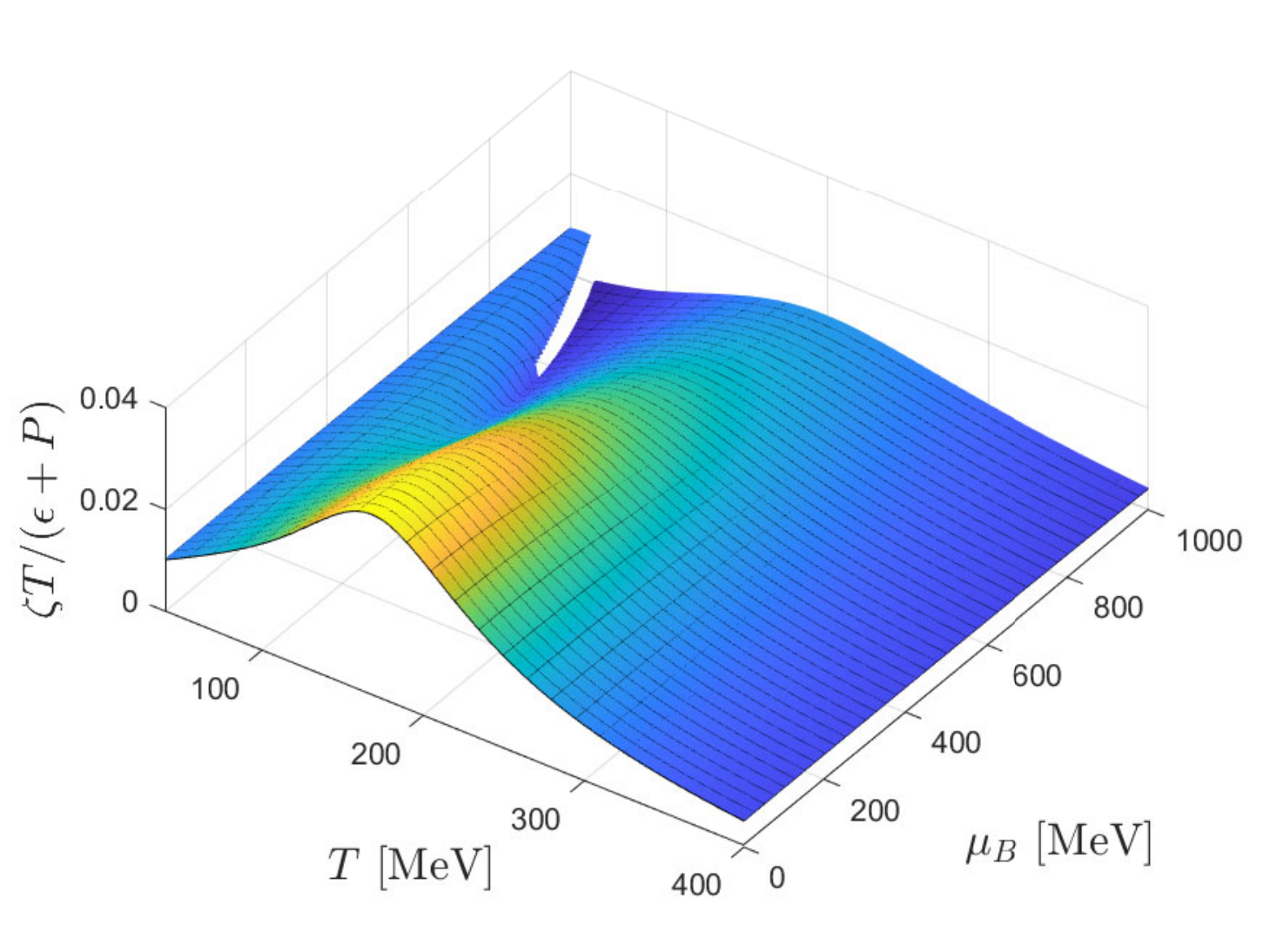}
    \includegraphics[width=0.5\textwidth]{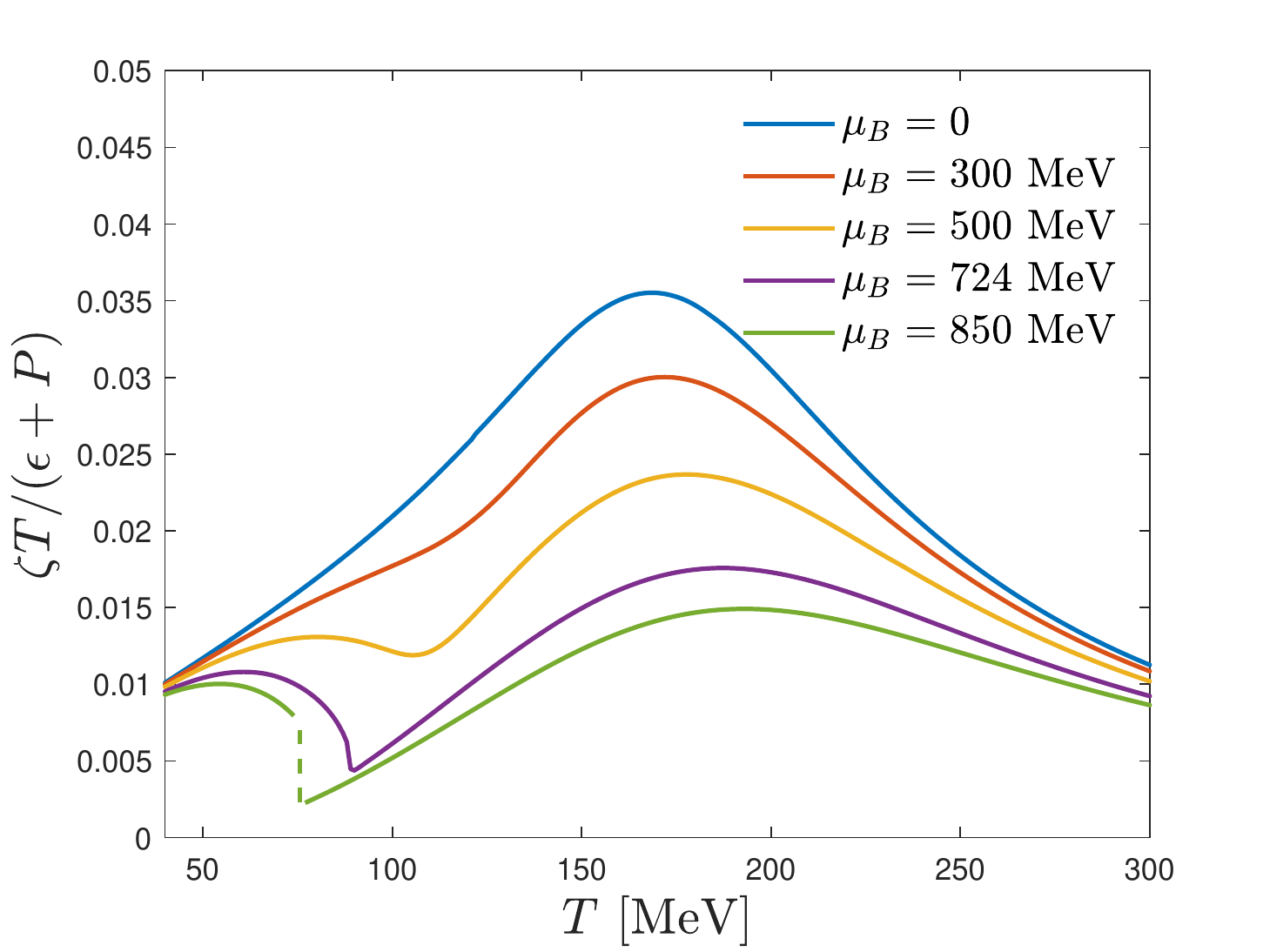}
    \caption{Holographic bulk viscosity $\zeta T/(\epsilon+P)$ as a function of $T$ and $\mu_B$ (top), and the same observable as a function of the temperature at different constant values of $\mu_B$ (bottom).}
    \label{fig:bulk_viscosity}
\end{figure}

The bulk viscosity $\zeta$, which measures the resistance of the fluid to deformation due to compression or expansion, in this holographic context can be associated with the gauge and diffeomorphism invariant combination of the EMD fields transforming as a singlet under $SO(3)$. The derivation of the differential equation associated with this calculation can be found in Ref. \cite{DeWolfe:2011ts}, and the details regarding the numerical computation when considering the present EMD model are reported in Ref. \cite{Grefa:2022sav}. The results regarding the bulk viscosity, presented as a dimensionless combination when divided by the enthalpy density are shown in Fig. \ref{fig:bulk_viscosity}. 

Interestingly, the bulk viscosity is suppressed with increasing baryon chemical potential indicating the fluid becomes even more ideal at higher densities. Also, the peak the bulk viscosity exhibits at vanishing chemical potential does not follow the CEP when $\mu_{B}$ increases. However, at some finite value of baryon chemical potential, the bulk viscosity develops an inflection point that gives rise to a local minimum at a larger value of $\mu_{B}$. Both, the inflection line and the local minimum of the bulk viscosity, follow and meet at the critical point. At the CEP, the holographic $\zeta$ remains finite, in contrast to the prediction for this observable in other effective models regarding a divergent bulk viscosity \cite{Monnai:2016kud,Buchel:2009mf}
although it exhibits a divergent slope. This is possibly due to the different dynamical universality class expected for QCD (type H)  \cite{Son:2004iv}, and the one for large-$N_{c}$/holographic approaches (type B) \cite{Natsuume:2010bs}.

\subsection{Shear Viscosity}

The shear viscosity measures the fluid resistance to shear flow in the presence of a velocity gradient within the layers of the fluid. In the holographic context, not only for the present model, but also any holographic model with isotropic, rotationally and translationally invariant holographic backgrounds with at most two derivatives in the gravity action, the holographic shear viscosity ratio is given by $\eta/s=1/4\pi$. 

\begin{figure}
    \centering
    \includegraphics[width=0.5\textwidth]{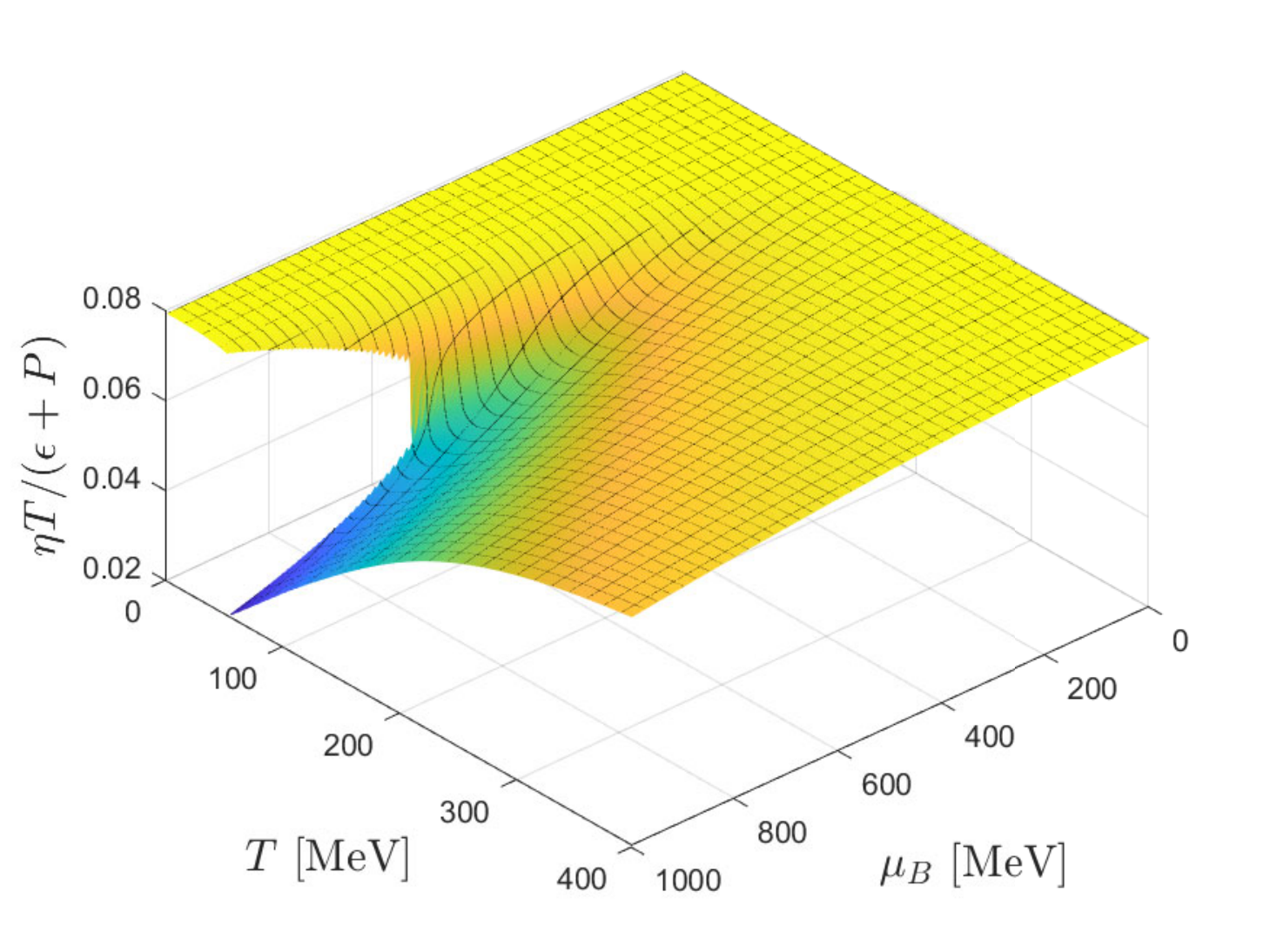}
    \includegraphics[width=0.5\textwidth]{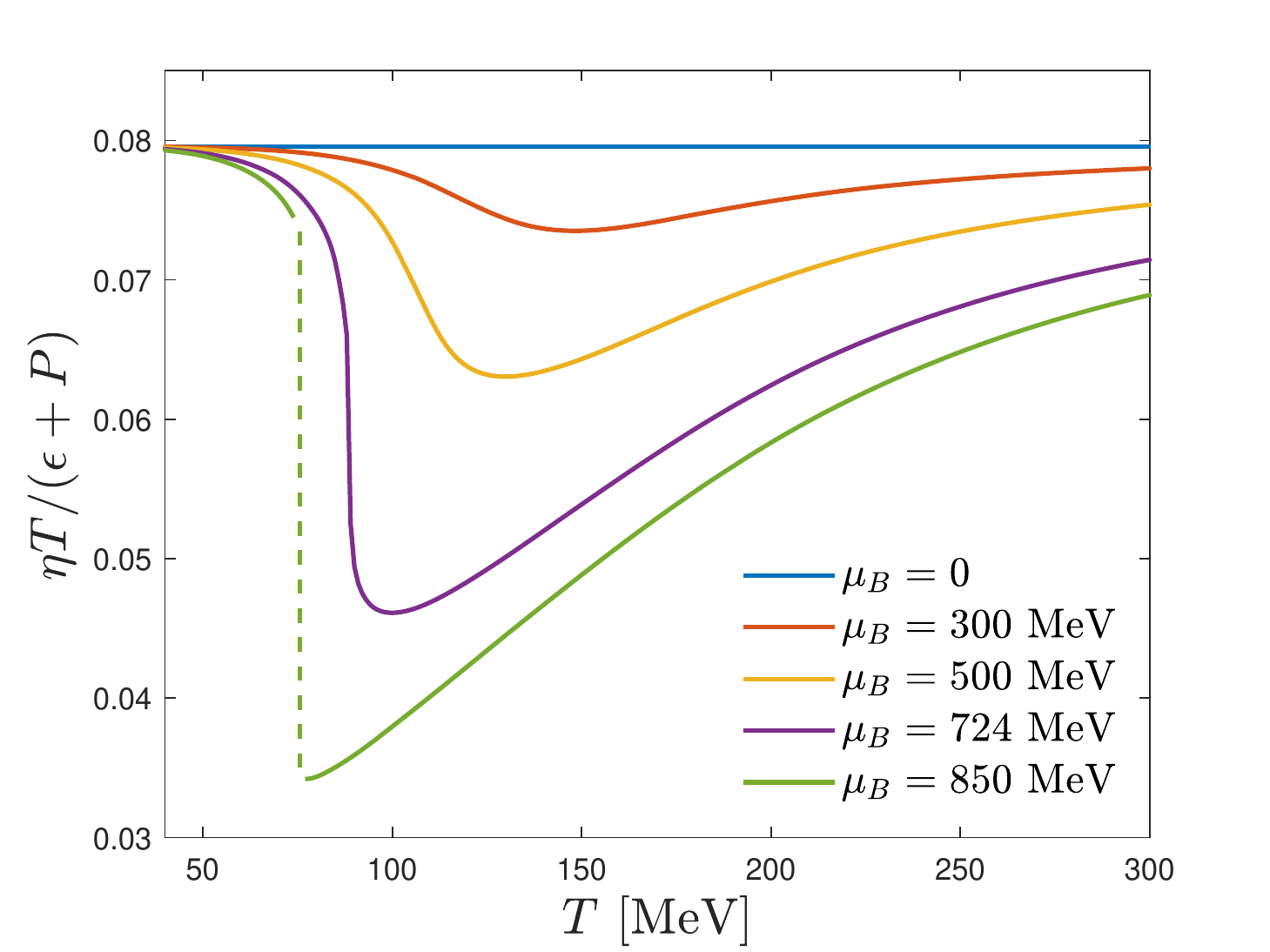}
    \caption{Holographic shear viscosity times temperature over enthalpy density as a function of $T$ and $\mu_B$ (top) and the same observable as a function of the temperature at several constant values of $\mu_B$ (bottom). This dimensionless combination reduces to $\eta/s=1/4\pi$ at $\mu_{B}=0$.}
    \label{fig:shear_visc}
\end{figure}

Similarly to the case for $\zeta$, the shear viscosity normalized by the enthalpy density represents the actual measure of fluidity in a baryon dense medium and is the relevant quantity entering in hydrodynamical simulations\cite{Denicol:2013nua,Liao:2009gb}. This dimensionless quantity reduces to the ratio $\eta/s$ at $\mu_{B}=0$. Also, the normalized holographic shear viscosity is suppressed with increasing $\mu_{B}$ and remains finite at the CEP, where it exhibits an infinite slope. The results for the normalized shear viscosity are presented in Fig. \ref{fig:shear_visc}.

\section{Conclusions}

Our holographic EMD model \cite{Critelli:2017oub,Grefa:2021qvt,Grefa:2022sav}, which is fixed to mimic the Lattice EoS for $\mu_{B}=0$, predicts a CEP at $T^{CEP}=89$ MeV and $\mu_{B}^{CEP}=724$ MeV. This model is in quantitative agreement with state-of-the-art
lattice QCD thermodynamics with (2 + 1) flavors and physical quark masses, both at zero and finite baryon density \cite{Critelli:2017oub,Grefa:2021qvt}.
With the first order phase transition line located, we considerably extended the baryon chemical potential coverage of the EoS in the phase diagram.
The transport coefficients related to transport of baryon charge, and viscosities were obtained over finite baryon chemical potential and particularly over the predicted line of first order phase transition and on top of the critical end point. We found that the holographic shear and bulk viscosities are suppressed with increasing baryon chemical potential, indicating the medium becomes closer to perfect fluidity at larger density. Additionally, the baryon conductivity, and bulk and shear viscosities are finite at the CEP, although they exhibit a divergent slope at this location. Beyond the critical point, and similarly to the equilibrium variables, the transport coefficients exhibit a discontinuity that corresponds to the line of first order phase transition. 

\section*{ACKNOWLEDGMENTS}
\vspace{-0.4cm}
This material is based upon work supported by the National Science Foundation under grants no. PHY-1654219 and PHY-2116686. This work was supported in part by the National Science Foundation (NSF) within the framework of the MUSES collaboration, under grant number OAC-2103680, the US-DOE Nuclear Science Grant No. DE-SC0020633, US-DOE Office of Science, Office of Nuclear Physics, within the framework of the Beam Energy Scan Topical (BEST) Collaboration. J.N. is partially supported by the U.S. Department of Energy, Office of Science, Office for Nuclear Physics under Award No. DE-SC0021301.

\bibliography{references}

\end{document}